# FINANCIAL VARIABLES EFFECT ON THE U.S. GROSS PRIVATE DOMESTIC INVESTMENT (GPDI) 1959-2001

## Byron E. Bell

Department of Mathematics, Olive-Harvey College
Chicago, Illinois, 60628, USA

# Abstract

I studied what role the US stock markets and money markets have possibly played in the Gross Private Domestic Investment (GPDI) of the United States from the year 1959 to the year 2001, Gross Private Domestic Investment refers to the total amount of investment spending by businesses and firms located within the borders of a nation. It includes both the values of the purchases of non-residential fixed investment, which include capital goods used for production, and the values of the purchases of residential fixed investment, which include construction spending for factories or offices. And I created a Multiple Linear Regression Model of the GDPI. To see if companies and private citizens use the stock market and money markets as a way of financing capital projects (business ventures, buying commercial and noncommercial property, etc).

## 0. INTRODUCTION

I will in this paper examine the mathematical statistical relationship between U.S. Gross Private Domestic Investment (GPDI) a dependent variable and the Dow Jones Industry of 30 Stocks (DJ), Standard and Poor's Index of 500 Stocks (SP500), New York Stock Exchange Index (NYSE), Consumer Price Index-Urban (CPI-U) and Three Month Treasury Bill's Rate (TB3) which are the independent variable(s) using data from the year 1959 to the year 2001 and carry out a regression analysis. Data for this study came from the Council of Economic Advisors (CEA), The Economic Report of the President (2003, 2002). In section 1 of this work; I will compare and use Pearson Correlation of stock indices.

In section 2; I will once again use Pearson Correlation of two (2) stock indices and CPI-U and produce a simple linear regression equation where the CPI-U is the dependent variable and SP500 is independent variable. In section 3; SP500, CPI-U is the independent variables and the dependent variable (TB3) will become a linear regression equation. In section 4; a multiple linear regression equation, model of the dependent variable will be form from the independent variables in the regression equation. An Analysis of Variances (ANOVA) table will be generated.

## 1. NYSE, SP500, DJ

I will use the following statistical theory (Pearson's product-moment coefficient) to show the relationship between NYSE and other variables of the stock market (SP500, DJ).

### New York Stock Exchange Index
### NYSE 1959–2001

Fig 1-1

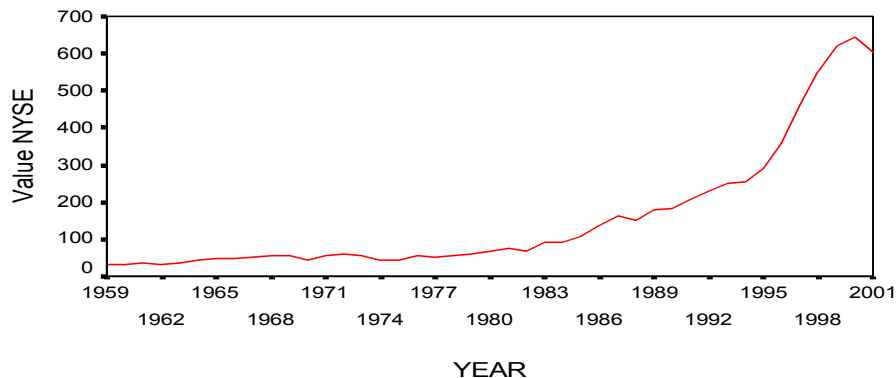

# Dow Jones Industrial Average 1959–2001

## DJ

### Fig 1-2

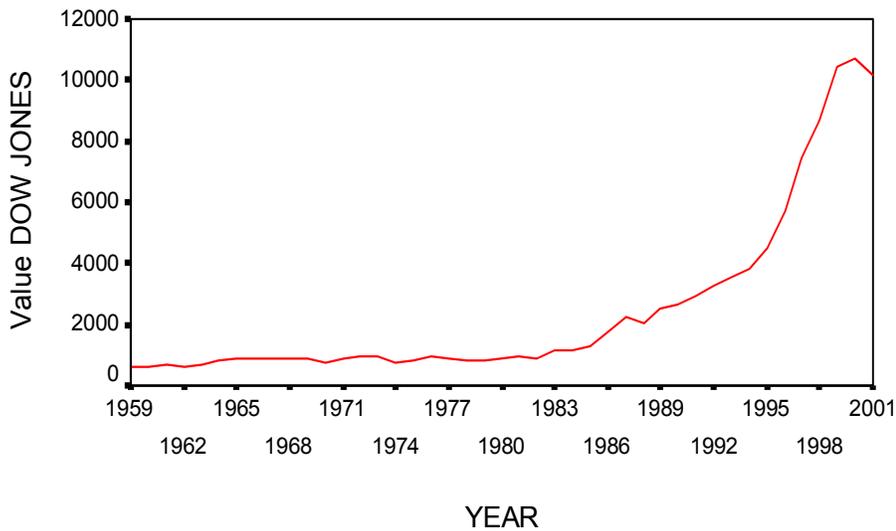

YEAR

# Standard & Poor 500 (SP 500) 1959–2001

### Fig 1-3

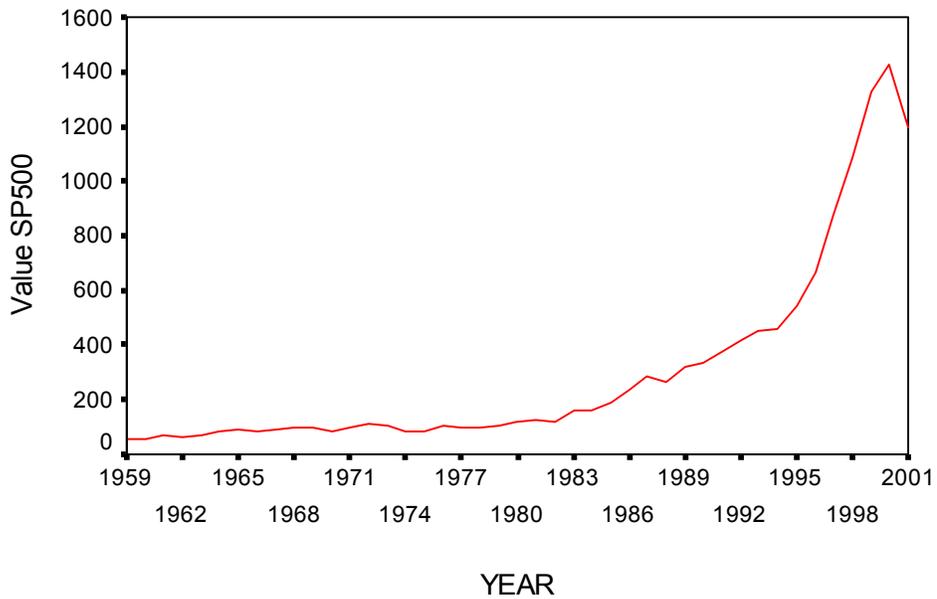

YEAR

## SCATTERPLOT OF NYSE AND DJ

### Fig 1-4

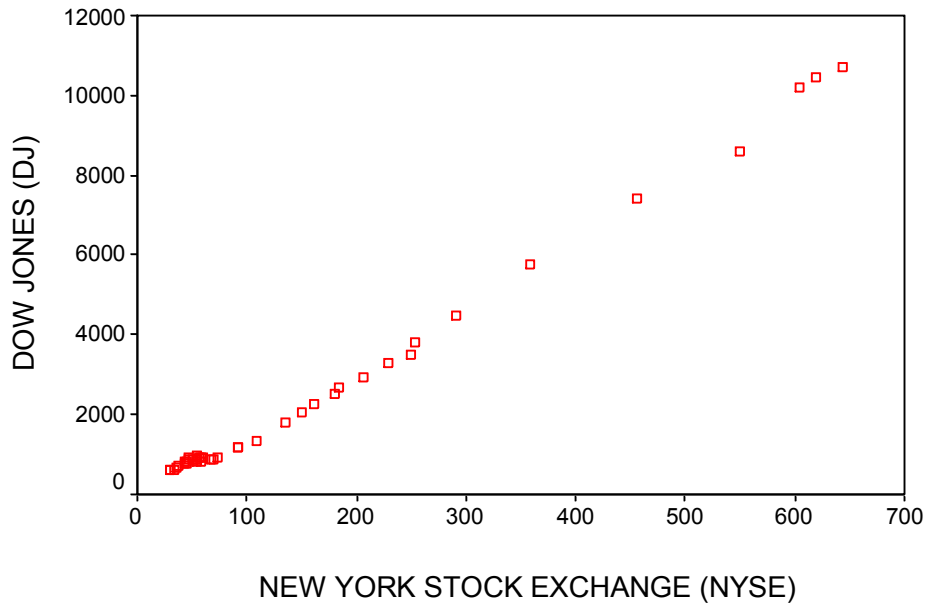

## SCATTERPLOT OF SP500 AND DJ

### Fig 1-5

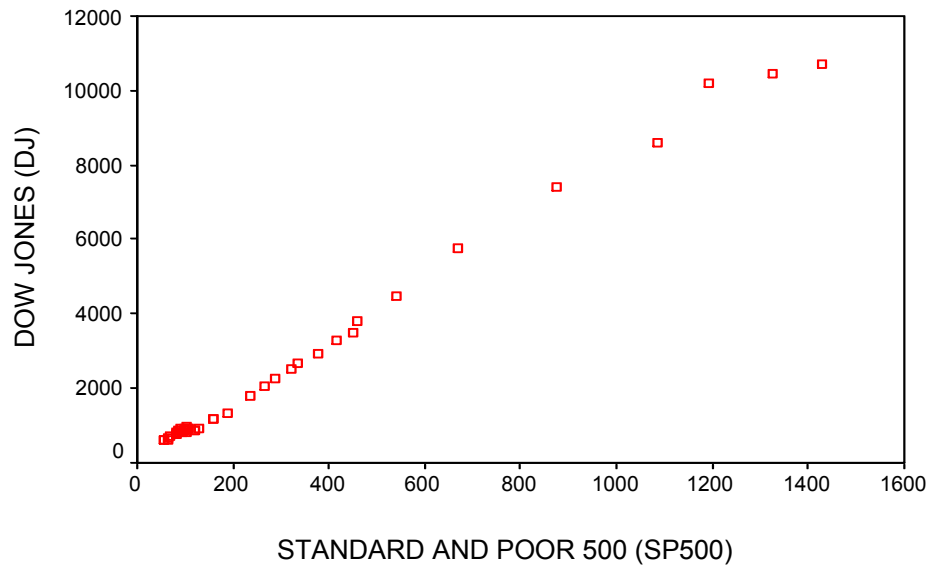

# SCATTERPLOT OF NYSE AND SP500

## Fig 1-6

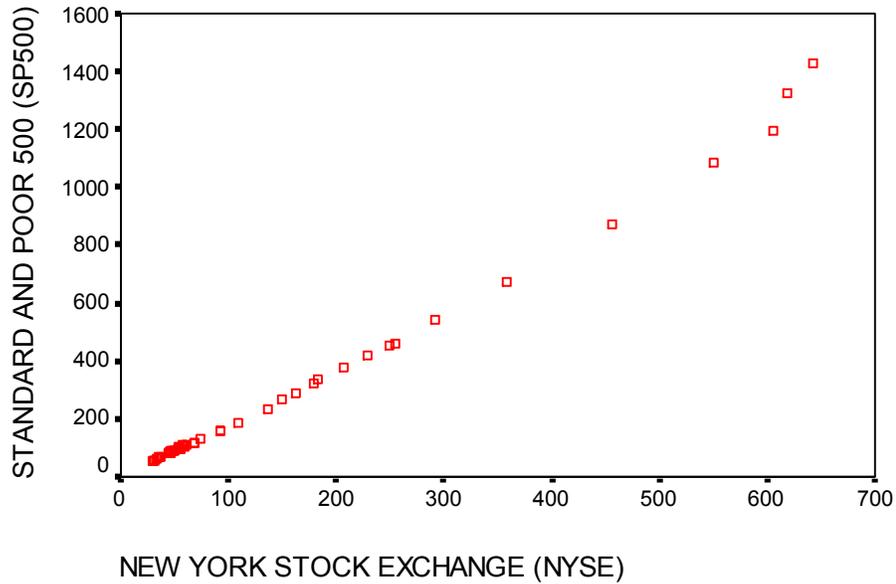

## Table 1-1

**Correlations**

|  |  | NEW YORK STOCK EXCHANGE (NYSE) | DOW JONES (DJ) | STANDARD AND POOR 500 (SP500) |
|---|---|---|---|---|
| NEW YORK STOCK EXCHANGE (NYSE) | Pearson Correlation | 1 | .997** | .996** |
|  | Sig. (2-tailed) | . | .000 | .000 |
|  | Sum of Squares and Cross-products | 1269247.942 | 20798852.1 | 2620505.307 |
|  | Covariance | 30220.189 | 495210.763 | 62392.983 |
|  | N | 43 | 43 | 43 |
| DOW JONES (DJ) | Pearson Correlation | .997** | 1 | .998** |
|  | Sig. (2-tailed) | .000 | . | .000 |
|  | Sum of Squares and Cross-products | 20798852.052 | 343161317 | 43142191.72 |
|  | Covariance | 495210.763 | 8170507.553 | 1027195.041 |
|  | N | 43 | 43 | 43 |
| STANDARD AND POOR 500 (SP500) | Pearson Correlation | .996** | .998** | 1 |
|  | Sig. (2-tailed) | .000 | .000 | . |
|  | Sum of Squares and Cross-products | 2620505.307 | 43142191.7 | 5449048.019 |
|  | Covariance | 62392.983 | 1027195.041 | 129739.239 |
|  | N | 43 | 43 | 43 |

**. Correlation is significant at the 0.01 level (2-tailed).

In examining the data of the above variables in we can see that the DJ index and SP500 index has the highest Pearson Correlation (.998) of the three variables. Pearson Correlation of the two variables (DJ index and SP500 index) is significant at the 0.01 level (2-tailed).

## 2. CPI-U

In this section, I am carrying out further analysis of the data from section 1, I will analyze the two indexes the DJ index and SP500 index which has the highest Pearson Correlation (.998) of the three stock indexes. A Pearson Correlation run of comparing the two indexes the DJ index and SP500 index and Consumer Price Index-Urban Area's (CPI-U) will be done. The stock index with the highest Pearson Correlation with CPI-U a simple linear regression equation will be made where the stock index is the independent variable and CPI-U is the dependent variable.

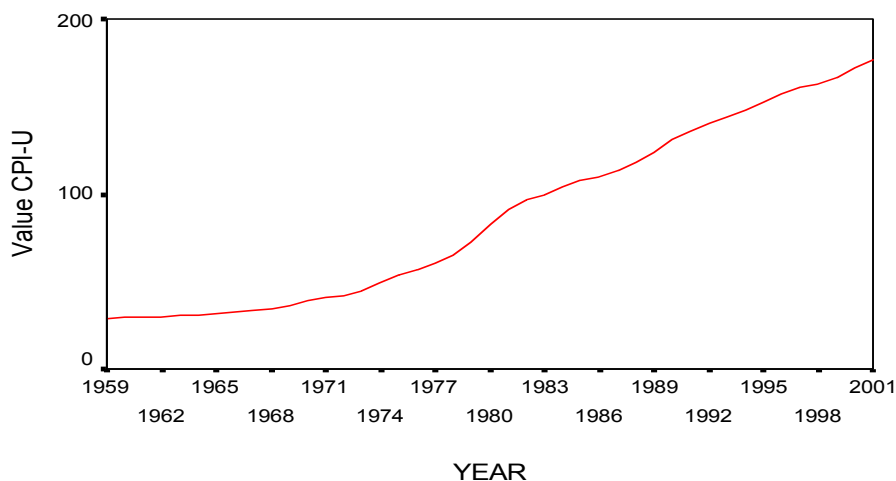

Consumer Price Index (CPI-U) 1959–2001

Fig 2-1

## SCATTERPLOT OF CPI-U and SP500

## Fig 2-2

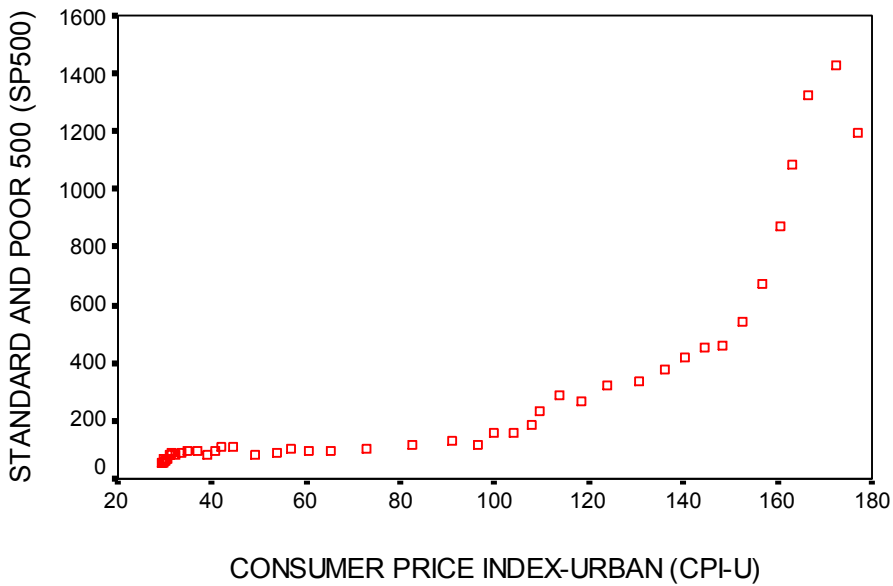

### Correlations
### Table 2-1

**Correlations**

|  |  | DOW JONES (DJI) | CPI-U |
|---|---|---|---|
| DOW JONES (DJ) | Pearson Correlation | 1 | .811** |
|  | Sig. (2-tailed) | . | .000 |
|  | N | 43 | 43 |
| CPI-U | Pearson Correlation | .811** | 1 |
|  | Sig. (2-tailed) | .000 | . |
|  | N | 43 | 43 |

**. Correlation is significant at the 0.01 level (2-tailed).

## Table 2-2

**Correlations**

|  |  | CPI-U | STANDARD AND POOR 500 (SP500) |
|---|---|---|---|
| CPI-U | Pearson Correlation | 1 | .820** |
|  | Sig. (2-tailed) | . | .000 |
|  | N | 43 | 43 |
| STANDARD AND POOR 500 (SP500) | Pearson Correlation | .820** | 1 |
|  | Sig. (2-tailed) | .000 | . |
|  | N | 43 | 43 |

**. Correlation is significant at the 0.01 level (2-tailed).

The variables that are correlated the highest with CPI-U the most is the SP500. The Pearson Correlation of the above two variables is .82. From this I will make a linear regression equation of SP500 as the independent variable and the CPI-U as the dependent variable.

## Regression

### Table 2-3

**Model Summary**

| Model | R | R Square | Adjusted R Square | Std. Error of the Estimate |
|---|---|---|---|---|
| 1 | .820a | .672 | .664 | 29.3379 |

a. Predictors: (Constant), STANDARD AND POOR 500 (SP500)

### Table 2-4

**ANOVAb**

| Model |  | Sum of Squares | df | Mean Square | F | Sig. |
|---|---|---|---|---|---|---|
| 1 | Regression | 72326.044 | 1 | 72326.044 | 84.030 | .000a |
|  | Residual | 35289.246 | 41 | 860.713 |  |  |
|  | Total | 107615.3 | 42 |  |  |  |

a. Predictors: (Constant), STANDARD AND POOR 500 (SP500)

b. Dependent Variable: CONSUMER PRICE INDEX-URBAN (CPI-U)

## Table 2-5

**Coefficients**[a]

| Model | | Unstandardized Coefficients | | Standardized Coefficients | | |
|---|---|---|---|---|---|---|
| | | B | Std. Error | Beta | t | Sig. |
| 1 | (Constant) | 52.892 | 5.863 | | 9.021 | .000 |
| | STANDARD AND POOR 500 (SP500) | .115 | .013 | .820 | 9.167 | .000 |

a. Dependent Variable: CONSUMER PRICE INDEX-URBAN (CPI-U)

The equation of Table 2-5 is CPI-U=52.892+.115*SP500          (1)

(.013)

In Table 2.3, R=.82, R-Square=.672, Adj R-Square=.664.

The above equation and tables the R-Square indicator tells us that the closer to 1 the more that the independent variable is related to the dependent variable. If the R-Square is closer to zero the there be little to no relationship between the independent variable (SP500) and the dependent variable (CPI-U) from Table 2.3.

The following Hypothesis T-test is base on Table 2.4:

Hypothesis T-test (Two-tailed test)

$H_0: \rho = 0$

$H_1: \rho \neq 0$

t=9.167, Alpha=.05, Sig. of SP500=.000

Alpha >Sig.→ Reject $H_0$

Alpha < Sig.→ Accept $H_0$

.05>.000→ Reject $H_0$

The Coefficients Table, Table 2-6 contains for each of the regression coefficients, their Standard Error ( Std. Error) is the same as Standard Deviations, as well as the t-ratios and p-values for testing the hypothesis that a coefficient is zero ( the variable has no significant effect on the dependent variable). The p-value or (Significant) Sig of .000 for SP500 in the Coefficients Table indicates that there is significant evidence of a nonzero population slope. The decision is to reject $H_0$ at the Alpha=.05 level. Therefore it is a   statistically significant relationship between the SP500 Index and CPI-U.

## 3. TB3

In this section, I will make a multiple linear regression equation of two independent variables and a dependent variable TB3. An ANOVA table will be generated of the regression equation,

$$TB3=b_0+b_1*SP500+b_2*(CPI-U) \qquad (2).$$

## US T-Bills 3-Mo Rate (TB3) 1959–2001

### Fig. 3-1

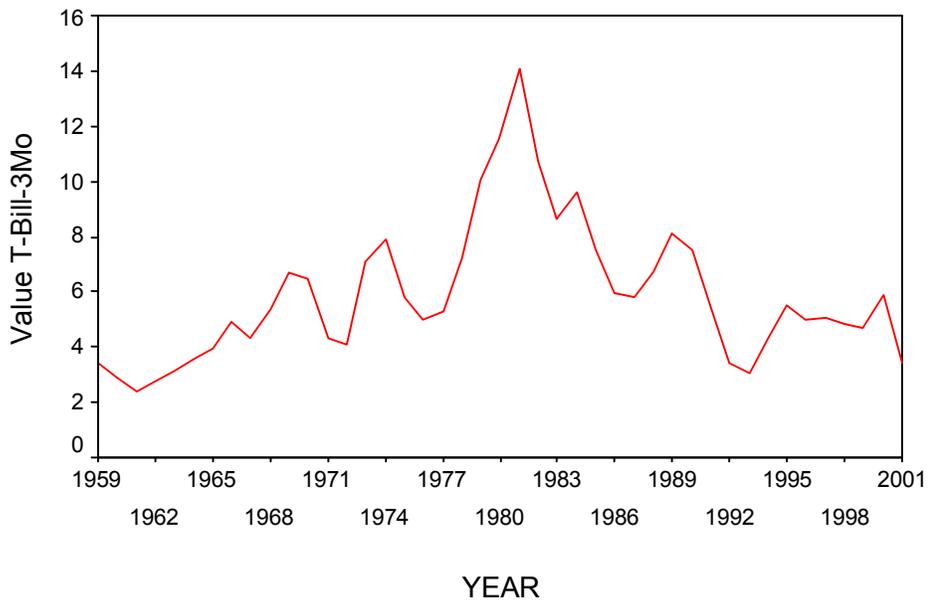

## SCATTERPLOT OF SP500 & TB3 , 1959–2001

### Fig. 3-2

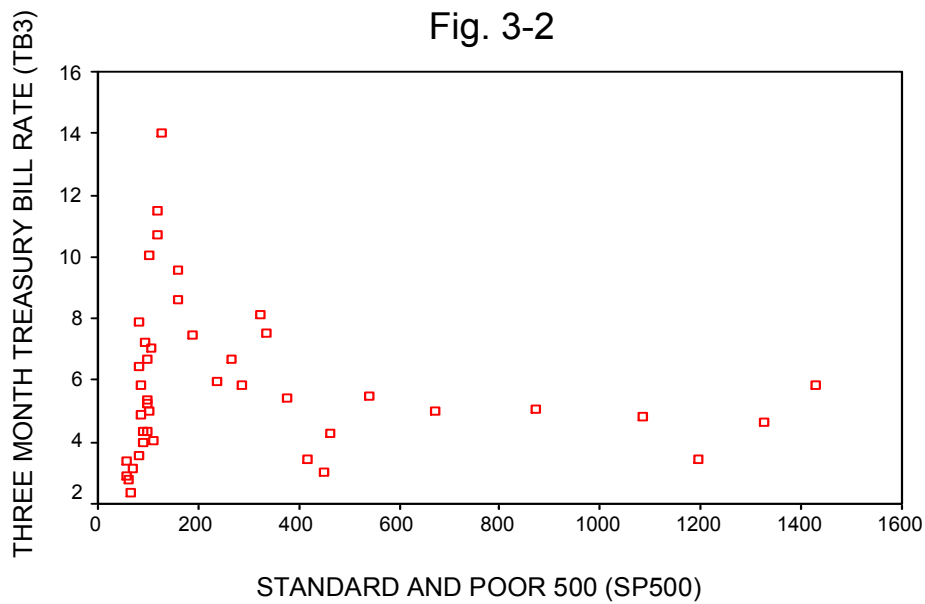

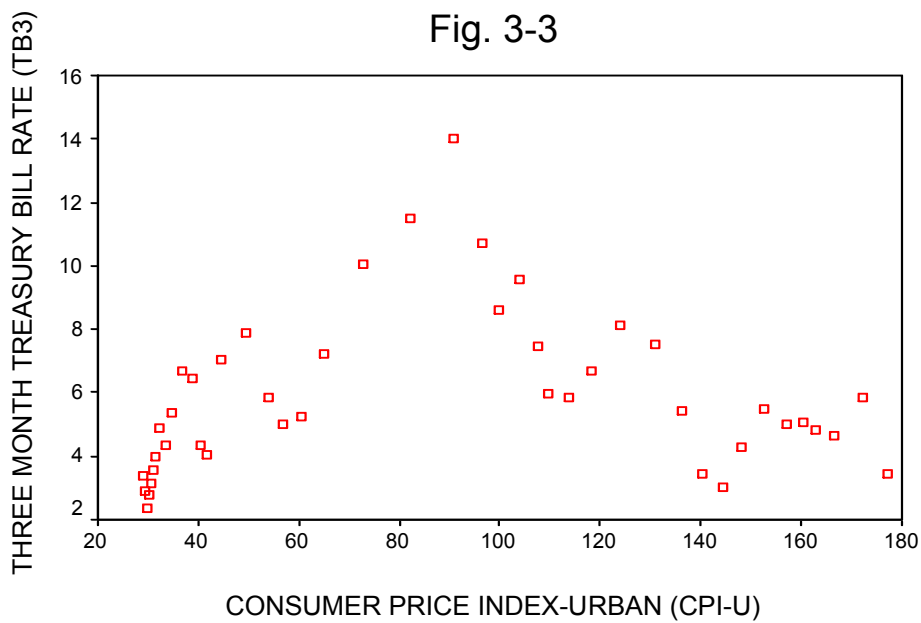

SCATTERPLOT OF CPI-U & TB3, 1959–2001

Fig. 3-3

**Regression**

**Table 3-1**

**Model Summary[b]**

| Model | R | R Square | Adjusted R Square | Std. Error of the Estimate |
|---|---|---|---|---|
| 1 | .458[a] | .210 | .170 | 2.307538 |

a. Predictors: (Constant), CONSUMER PRICE INDEX-URBAN (CPI-U), STANDARD AND POOR 500 (SP500)

b. Dependent Variable: THREE MONTH TREASURY BILL RATE (TB3)

## Table 3-2

**ANOVA[b]**

| Model | | Sum of Squares | df | Mean Square | F | Sig. |
|---|---|---|---|---|---|---|
| 1 | Regression | 56.470 | 2 | 28.235 | 5.303 | .009[a] |
| | Residual | 212.989 | 40 | 5.325 | | |
| | Total | 269.459 | 42 | | | |

a. Predictors: (Constant), CONSUMER PRICE INDEX-URBAN (CPI-U), STANDARD AND POOR 500 (SP500)

b. Dependent Variable: THREE MONTH TREASURY BILL RATE (TB3)

## Table 3-3

**Coefficients[a]**

| Model | | Unstandardized Coefficients | | Standardized Coefficients | t | Sig. |
|---|---|---|---|---|---|---|
| | | B | Std. Error | Beta | | |
| 1 | (Constant) | 4.278 | .797 | | 5.370 | .000 |
| | STANDARD AND POOR 500 (SP500) | -.005 | .002 | -.778 | -3.170 | .003 |
| | CONSUMER PRICE INDEX-URBAN (CPI-U) | .037 | .012 | .743 | 3.026 | .004 |

a. Dependent Variable: THREE MONTH TREASURY BILL RATE (TB3)

In Table 3-1, R=.458, R-Square=.210, Adj R-Square=.170.

The equation of Table 3-3 is
$$TB3 = 4.278 - .005*SP500 + .037*CPI\text{-}U \qquad (3)$$
$$(.002) \qquad (.012)$$

From Table 3-2

### Hypothesis F-test

$H_0$: $\rho=0$

$H_1$: $\rho\neq0$

F=5.303, Alpha=.05, Sig. of SP500 & CPI-U=.009

Alpha >Sig.$\rightarrow$ Reject $H_0$
Alpha < Sig.$\rightarrow$ Accept $H_0$

.05>.009$\rightarrow$ Reject $H_0$

from the above tables is computed an F-ratio to test that all of the independant variables coefficients are zero and prints the result in an ANOVA Table which is the above Table 3-2. In this model, the F-value of 5.303 corresponds to a p-value or (Significant) Sig of .009.

The decision is to reject $H_0$ at the Alpha=.05 level. Therefore it is a statistically significant relationship between the dependant variable TB3 and independent variables of SP500 & CPI-U.

## 4. U.S. GPDI

In this section, I will make a multiple linear regression equation of three independent variables and a dependent variable GPDI. An ANOVA table will be generated that will give clearer analysis of the regression equation,

$$GPDI=b_0+b_1*SP500+b_2*(CPI-U) +b_3*TB3, \qquad (4)$$

## Gross Private Domestic Investment 1959–2001

### GPDI

### Fig 4-1

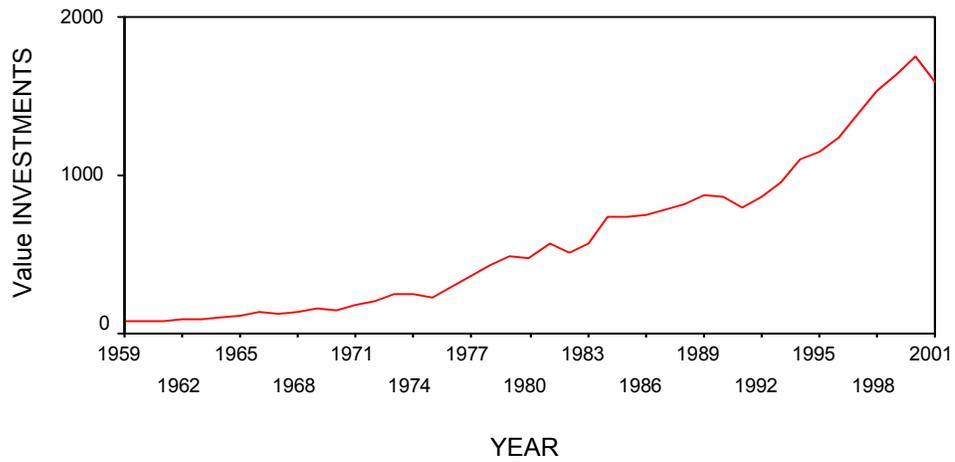

## SCATTERPLOT OF GPDI AND SP500

### Fig 4-2

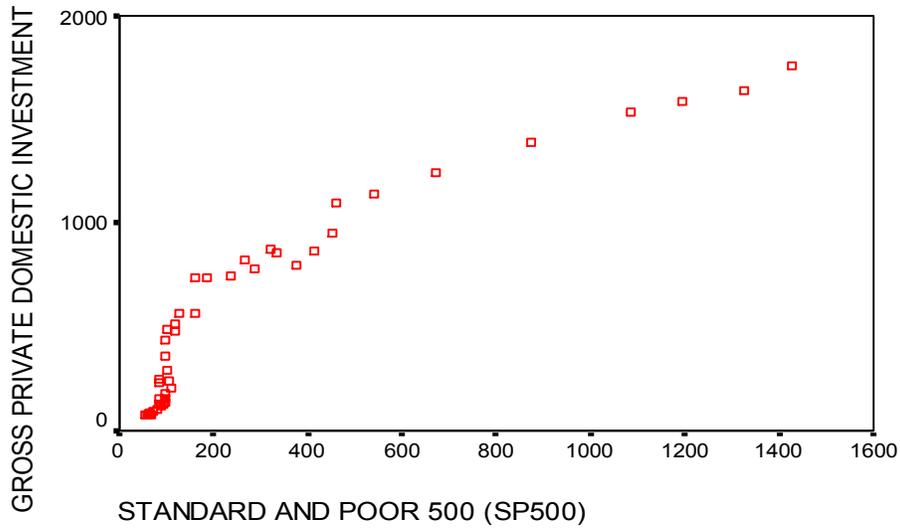

SCATTERPLOT OF GPDI AND CPI-U

Fig 4-3

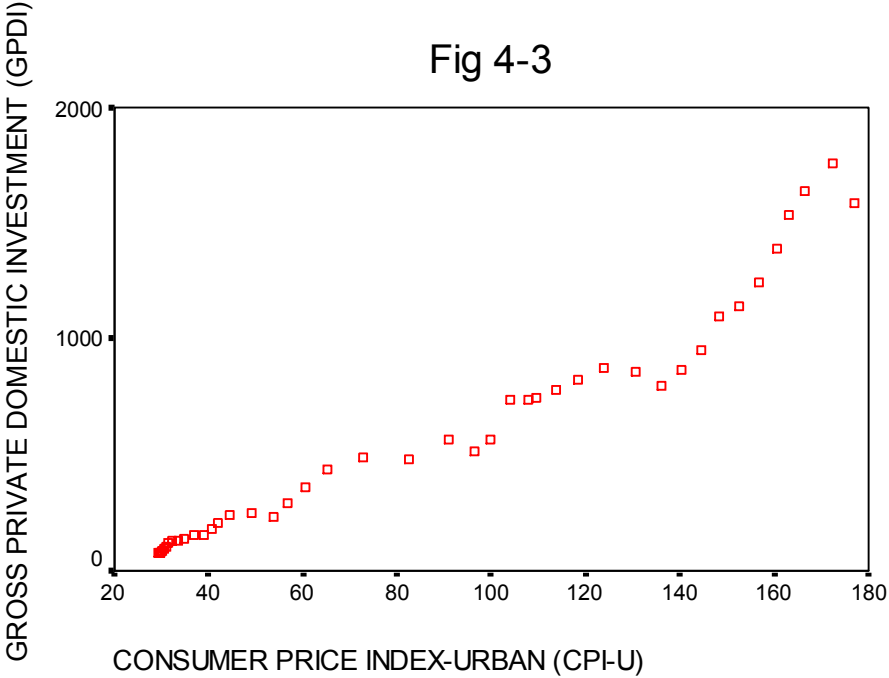

SCATTERPLOT OF GPDI AND TB3

Fig 4-4

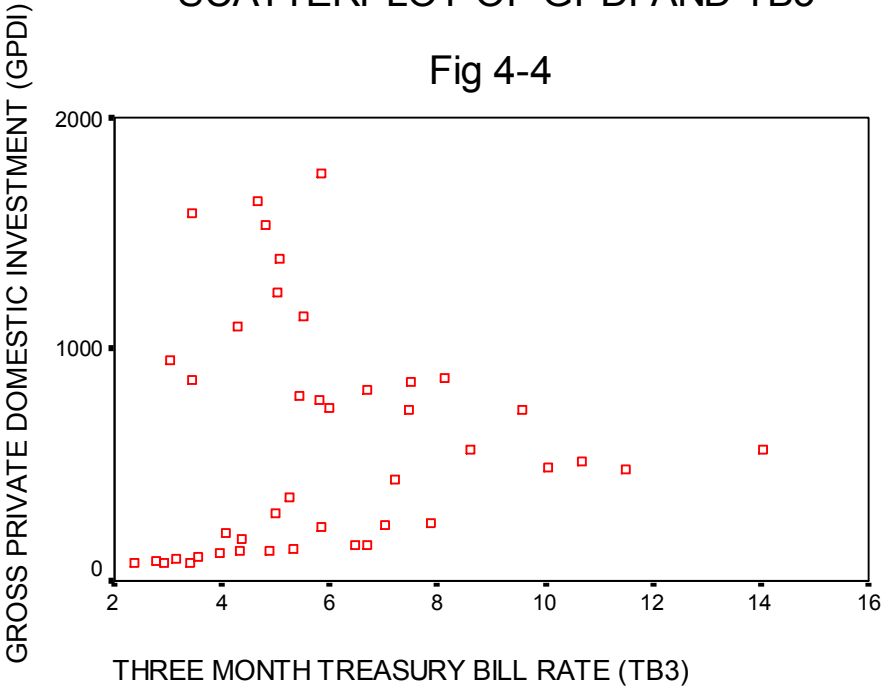

# Regression

## Table 4-1

**Model Summary**

| Model | R | R Square | Adjusted R Square | Std. Error of the Estimate |
|---|---|---|---|---|
| 1 | .996a | .992 | .991 | 46.0644 |

a. Predictors: (Constant), THREE MONTH TREASURY BILL RATE (TB3), CONSUMER PRICE INDEX-URBAN (CPI-U), STANDARD AND POOR 500 (SP500)

## Table 4-2

**ANOVAb**

| Model | | Sum of Squares | df | Mean Square | F | Sig. |
|---|---|---|---|---|---|---|
| 1 | Regression | 10077978 | 3 | 3359326.054 | 1583.148 | .000a |
| | Residual | 82755.177 | 39 | 2121.928 | | |
| | Total | 10160733 | 42 | | | |

a. Predictors: (Constant), THREE MONTH TREASURY BILL RATE (TB3), CONSUMER PRICE INDEX-URBAN (CPI-U), STANDARD AND POOR 500 (SP500)

b. Dependent Variable: GROSS PRIVATE DOMESTIC INVESTMENT (GPDI)

## Table 4-3

**Coefficientsa**

| Model | | Unstandardized Coefficients | | Standardized Coefficients | t | Sig. |
|---|---|---|---|---|---|---|
| | | B | Std. Error | Beta | | |
| 1 | (Constant) | -162.815 | 20.865 | | -7.803 | .000 |
| | STANDARD AND POOR 500 (SP500) | .574 | .039 | .420 | 14.890 | .000 |
| | CONSUMER PRICE INDEX-URBAN (CPI-U) | 6.031 | .272 | .621 | 22.188 | .000 |
| | THREE MONTH TREASURY BILL RATE (TB3) | 10.144 | 3.156 | .052 | 3.214 | .003 |

a. Dependent Variable: GROSS PRIVATE DOMESTIC INVESTMENT (GPDI)

In Table 4-1, R=.996, R-Square=.992, Adj R-Square=.991.

The equation of Table 4-3 is

$$GPDI=-162.815+.574*SP500+6.031*(CPI\text{-}U)+10.144*TB3 \qquad (5)$$
$$(.039) \qquad (.272) \qquad (3.156)$$

Hypothesis F-test of Table 4-2 is

Hypothesis F-test

$H_0{:}\ \rho=0$

$H_1{:}\rho\neq0$

F=1583.148, Sig. of SP500, CPI-U and TB3=.000

Alpha >Sig.$\rightarrow$ Reject $H_0$

Alpha < Sig.$\rightarrow$ Accept $H_0$

.05>.000$\rightarrow$ Reject $H_0$

from the tables above is computed an F-ratio to test that all of the

independant variables coefficients are zero and prints the result in an

ANOVA Table which is the above Table 4-2.

In this model, the F-value of 1583.148 corresponds to a p-value or

(Significant) Sig of .000. The decision is to reject $H_0$ at the Alpha=.05

level. Therefore it is a statistically significant relationship between the

dependant variable GPDI and independant variables of SP500, CPI-U

and TB3.

## 5. CONCLUSION

The above study should be looked at only as a possible trend model not a trading model of the stock market. Further investigations are needed to develop a trading model. Studies in nonlinear mathematics and modeling (Non-linear Statistics, Dynamic Theory) are needed and real world testing of the data to the relationship between theories and how the stock market reacts is a must. Research of other variables that effect stocks and interest rates should be done. For example, the CPI-U has the smallest Pearson Correlation in relationship to GPDI. Standardized Coefficients from the regression equation known as Beta and variances analysis testing are needed. The mathematical statistical methods employed in this current work are from Hogg, R.V., Tanis, E.A. (2001) and Hogg, R.V., Craig, A.T. (1965). All of this material is from my research in Bell, B.E. (2006). A correlation table and graph of all variables in this paper is included.

# APPENDIX

# Correlations

| | | GROSS PRIVATE DOMESTIC INVESTMENT (GPDI) | NEW YORK STOCK EXCHANGE (NYSE) | DOW JONES (DJ) | STANDARD AND POOR 500 (SP500) | CPI-U | T-BILL-3MO |
|---|---|---|---|---|---|---|---|
| GROSS PRIVATE DOMESTIC INVESTMENT (GPDI) | Pearson Correlation | 1 | .939(**) | .912(**) | .920(**) | .971(**) | .046 |
| | Sig. (2-tailed) | . | .000 | .000 | .000 | .000 | .768 |
| | N | 43 | 43 | 43 | 43 | 43 | 43 |
| NEW YORK STOCK EXCHANGE (NYSE) | Pearson Correlation | .939(**) | 1 | .997(**) | .996(**) | .852(**) | -.167 |
| | Sig. (2-tailed) | .000 | . | .000 | .000 | .000 | .283 |
| | N | 43 | 43 | 43 | 43 | 43 | 43 |
| DOW JONES (DJ) | Pearson Correlation | .912(**) | .997(**) | 1 | .998(**) | .811(**) | -.197 |
| | Sig. (2-tailed) | .000 | .000 | . | .000 | .000 | .206 |
| | N | 43 | 43 | 43 | 43 | 43 | 43 |
| STANDARD AND POOR 500 (SP500) | Pearson Correlation | .920(**) | .996(**) | .998(**) | 1 | .820(**) | -.169 |
| | Sig. (2-tailed) | .000 | .000 | .000 | . | .000 | .278 |
| | N | 43 | 43 | 43 | 43 | 43 | 43 |
| CPI-U | Pearson Correlation | .971(**) | .852(**) | .811(**) | .820(**) | 1 | .105 |
| | Sig. (2-tailed) | .000 | .000 | .000 | .000 | . | .503 |
| | N | 43 | 43 | 43 | 43 | 43 | 43 |
| T-BILL-3MO | Pearson Correlation | .046 | -.167 | -.197 | -.169 | .105 | 1 |
| | Sig. (2-tailed) | .768 | .283 | .206 | .278 | .503 | . |
| | N | 43 | 43 | 43 | 43 | 43 | 43 |

** Correlation is significant at the 0.01 level (2-tailed).

## Correlations
### Graph of all of the Variables in this project

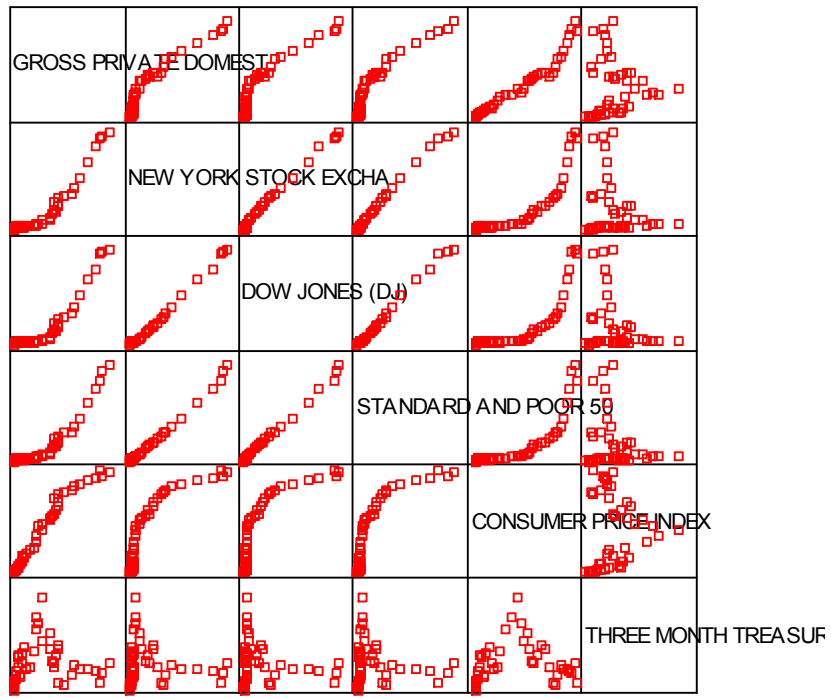

BYRON E. BELL
Department of Mathematics,
Olive-Harvey College
10001 S. Woodlawn
Chicago, Illinois, 60628, USA
bbell@ccc.edu